\documentclass[prb,twocolumn,flowatfix,showpacs,superscriptaddress]{revtex4}
\usepackage{graphicx}
\usepackage{amsmath} 
\usepackage{amssymb}
\usepackage{dsfont}
\usepackage{color}
\graphicspath{{EPS/}}

\newcommand{\begeq}{\begin{equation}}           
\newcommand{\eeq}{\end{equation}}
\newcommand{\begeqn}{\begin{eqnarray}}
\newcommand{\eeqn}{\end{eqnarray}}

\newcommand{\up}{\uparrow}
\newcommand{\down}{\downarrow}
\newcommand{\G}{\Gamma}
\newcommand{\s}{\sigma}
\newcommand{\w}{\omega}

\newcommand{\eps}{\varepsilon}

\newcommand{\GG}{\mathcal{G}}
\newcommand{\GGh}{\hat{\mathcal{G}}}

\begin{document}

\title{Interplay of mesoscopic and Kondo effects 
for transmission amplitude of few-level quantum dots}
\author{T.~Hecht}\affiliation{Physics Department, Arnold Sommerfeld Center for Theoretical Physics and Center for NanoScience, Ludwig-Maximilians-Universit\"{a}t M\"{u}nchen, Germany}
\author{A.~Weichselbaum}\affiliation{Physics Department, Arnold Sommerfeld Center for Theoretical Physics and Center for NanoScience, Ludwig-Maximilians-Universit\"{a}t M\"{u}nchen, Germany}
\author{Y.~Oreg}\affiliation{Department of Condensed Matter Physics, The Weizmann Institute of Science, Rehovot 76100, Israel}
\author{J.~von Delft}\affiliation{Physics Department, Arnold Sommerfeld Center for Theoretical Physics and Center for NanoScience, Ludwig-Maximilians-Universit\"{a}t M\"{u}nchen, Germany}
\date{\today}

\begin{abstract}
The magnitude and phase of the transmission amplitude of a multi-level quantum dot is calculated for the mesoscopic regime of  level spacing large compared to level width. The interplay between Kondo correlations and the influence by neighboring levels is discussed. As in the single-level case, the Kondo plateaus of magnitude and phase disappear with increasing temperature. At certain gate voltages, ``stationary'' points are found at which the transmission phase is independent of temperature. Depending on the mesoscopic parameters of the adjacent levels (like relative sign and magnitude of tunneling matrix elements), the stationary points are shifted to or repelled by the neighboring level. 
\end{abstract}
\pacs{73.23.Hk, 73.23.-b, 73.63.Kv, 73.40.Gk}

\maketitle

\section{Introduction}

In a remarkable series of
experiments,\cite{Yacoby1995,Schuster1997,Ji2000,Ji2002,Avinun-Kalish2005,Zaffalon2007}
the Heiblum group has analyzed the complex transmission
amplitude, $t_d = |t_d| e^{i \alpha}$, of a quantum dot embedded in an
Aharonov-Bohm ring. In particular, by analyzing the Aharonov-Bohm oscillations of the conductance of such a ring, the dependence of both the
magnitude and phase of the transmission amplitude, $|t_d|$ and
$\alpha$, were measured as a function of various parameters such as
gate voltage $V_g$ applied to the dot, temperature $T$, mean coupling
strength to the leads $\Gamma$, etc.

The first two experiments in this series,\cite{Yacoby1995,Schuster1997} dealt with large dots containing many
$(> 100)$ electrons. 
The experiment by Yacoby \emph{et al.}\cite{Yacoby1995} showed that coherent transport through a quantum dot is possible despite the presence of strong interactions. The next experiments by Schuster \emph{et al.}\cite{Schuster1997} generated tremendous interest because the
behavior of the transmission phase showed a surprisingly ``universal''
behavior as function of gate voltage: the phase experienced a series
of sudden jumps by $- \pi$ (phase lapses) between each pair of Coulomb
blockade peaks in the conductance through the dot. This contradicted
a naive expectation that the behavior of the transmission phase
should depend on microscopic details of the dot, such as the signs of
the matrix elements coupling a given level to the left or right lead.

Subsequent experiments by Ji \emph{et al.}, \cite{Ji2000,Ji2002}
performed on smaller dots containing tens of electrons, analyzed how
the occurrence of the Kondo effect influences the transmission
amplitude, and in particular its phase. For transmission at zero
temperature through a \emph{single} level, the Kondo effect
causes the magnitude of the transmission amplitude to exhibit (as
function of gate voltage) a plateau at the unitary limit $(|t_d|=1)$. For this regime it
had been predicted by Gerland \emph{et al.} \cite{Gerland2000} that
the phase should show a plateau at  $\alpha = \pi/2$, a result very different from the
universal behavior mentioned above.  While the experiments of
Ji \emph{et al.} did yield deviations from the universal phase
behavior, they did not verify the prediction of a $\pi/2$ Kondo
plateau in the phase.  With hindsight, the reason probably was that
the experiments did not realize the conditions assumed in the
calculations of Gerland \emph{et al.},\cite{Gerland2000} namely transport through only a
\emph{single} level.

Truly ``mesoscopic'' behavior for the phase was observed only rather
recently by Avinun-Kalish \emph{et al.},\cite{Avinun-Kalish2005} in
even smaller dots containing only a small $(< 10)$ number of
electrons. For these, the mean level spacing $\delta$ was
significantly larger than the average level width $\Gamma$, so that
for any given gate voltage, transport through the dot is typically
governed by the properties of only a single level, namely that closest
to the Fermi energies of the leads.  When the number of electrons was
increased beyond about 14, universal behavior for the phase was
recovered. Consequently, it was proposed
\cite{Avinun-Kalish2005,Silvestrov2000,Silvestrov2001,Golosov2006,Karrasch2007a,Karrasch2007b,Oreg2007} that the universal
behavior occurs whenever a quantum dot is large enough for that the ratio
$\delta/\Gamma$ is sufficiently small $(\simeq 1)$ that for any given
gate voltage, typically more than one level contributes to transport.

The latest paper in this series, by Zaffalon \emph{et
  al.},\cite{Zaffalon2007} studied the transmission phase through a
quantum dot in the ``deep mesoscopic'' regime $\delta/\Gamma \gg 1$,
containing only one or two electrons.  When this system was tuned into
the Kondo regime, the transmission phase indeed did show the $\pi/2$
Kondo plateau predicted by Gerland \emph{et al.}\cite{Gerland2000}.

The experiments of Avinun-Kalish \emph{et al.},\cite{Avinun-Kalish2005} which observed
mesoscopic effects for the transmission phase through a small number
of levels, and those of Zaffalon \emph{et al.},\cite{Zaffalon2007} which found
characteristic signatures of the Kondo effect in the transmission
phase through a single level, raise the following question: what type
of phase behavior can arise in the deep mesoscopic regime from the
interplay of (i) \emph{random signs} for tunneling amplitudes of
neighboring levels and (ii) the \emph{Kondo effect} for individual levels?
In the present paper, we address this question by studying spin-degenerate models of
dots with 2 or 3 levels 
in the deep mesoscopic regime of $\delta / \Gamma \gg 1$. This is the regime relevant for the experiments of Zaffalon \emph{et al.}\cite{Zaffalon2007} (for those of Ji \emph{et al.},\cite{Ji2000,Ji2002} the ratio $\delta/\Gamma$ was presumably smaller than used here). Our goal is to provide a catalogue of
the types of behavior that can occur in this regime, and to illustrate
how the characteristic transmission amplitude (magnitude and phase)
depends on temperate as well as on the strength of the coupling to the leads.

This paper is organized as follows. In Sec.\ \ref{sec:model} we introduce our many-level model for the quantum dot system. We discuss the relation between the Aharonov-Bohm contribution to the linear conductance and the transmission amplitude through the quantum dot. The latter can be expressed in terms of the local Green's function of the dot. We briefly present the technique used to calculate the latter, the numerical renormalization group method.
In Sec.\ \ref{sec:Results} we present our numerical results of both the phase and the magnitude of the transmission amplitude through a two- and three-level model in the regime $\delta/\Gamma\gg 1$. We discuss the $T$- and $\Gamma$-dependence of the transmission amplitude with focus on the influence on Kondo correlations. We study all relevant choices of the mesoscopic parameters given by the relative signs of the tunneling amplitudes of adjacent levels. 
The influence of neighboring levels is studied. It results not only in a phase lapse in Coulomb blockade valleys but also introduces a $V_g$-asymmetry in the finite temperature modulations of the Kondo plateaus. ``Stationary'' points of $T$- and $\Gamma$-independence are discussed.
In the Appendix, we give a derivation of a formula for the Aharonov-Bohm contribution to the linear conductance through a multi-terminal interferometer with open geometry, as used in the Heiblum group. 
This formula has been used in several publications including some of the present authors, \cite{Gerland2000,Karrasch2007a,Karrasch2007b} but its derivation had not been published before.

\section{The model and the method}
\label{sec:model}

In the experiments,\cite{Schuster1997,Ji2000,Ji2002,Avinun-Kalish2005,Zaffalon2007} the temperature-dependent transmission amplitude through the
quantum dot is extracted from the Aharonov-Bohm oscillations of the
conductance in a multi-lead ring geometry. In the Appendix we show that this transmission amplitude can be
expressed in terms of the equilibrium local Green's function of the dot tunnel-coupled only to \emph{two} leads on its left and right side, without explicitly incorporating the other leads of the ring geometry in the calculation.

In this Section we introduce a ``reduced model'' describing the latter situation of a spinful multi-level quantum dot coupled to two reservoirs and present the transmission formula derived in the Appendix.
Further, we comment on NRG, the  method used to calculate the local Green's function.

\subsubsection{The model Hamiltonian}

The model Hamiltonian can be split into three parts,
\begin{subequations}
\begeq
	H=
	H_{\rm d}+H_{\rm l}+H_{\rm t},
	\label{AM:H}
\eeq
specifying the properties of the bare dot, the leads and the coupling between the two systems, respectively. 
For $N$ spinful levels coupled to a left (emitter) and right (collector) lead, these terms are given by
\begeqn
   H_{\rm d}
   &=& \sum_{j=1..N}\sum_{\sigma} \varepsilon_{dj} n_{dj\sigma} + 
   \hspace{-10pt}\sum_{\{j\sigma\}\neq \{j'\sigma'\}}\hspace{-10pt} U n_{dj\sigma} n_{dj'\sigma'} 
   \label{AM:H_imp}
   \\
   H_{\rm l}
   &=& \sum_{\alpha=L,R}\sum_{ k \s}\varepsilon_{ k} c^\dag_{\alpha  k \s}c_{\alpha  k \s}
   \label{AM:H_bath}
   \\	
   H_{\rm t}
   &=& \sum_{j} \sum_{\alpha=L,R} \sum_{k \s} 
   	(t^j_{\alpha} c^\dag_{\alpha k \s}d_{j\s} + 
   	\mbox{H.c.})\ .
   \label{AM:H_imp-bath}
\eeqn
\label{AM:Htot}
\end{subequations}
Dot creation operators for level $j$ and spin $\sigma\hspace{-3pt}=\hspace{-3pt}\{\up,\down\}$ are denoted by $d^\dag_{j\s}$, with $n_{dj\s}=d^\dag_{j\s}d_{j\s}$, where $j=1\cdots N$ labels the levels in order of increasing energy ($\varepsilon_{dj}<\varepsilon_{dj+1}$). We use an inter- and intra-level independent Coulomb energy $U>0$. The leads are assumed to be identical and non-interacting with constant density of states $\rho=1/2D$, where the half-bandwidth $D=1$ serves as energy unit. Electrons in lead $\alpha$ are created by $c_{\alpha k \s}^\dag$. The local levels are tunnel-coupled to the leads, with real overlap matrix elements $t^{j}_{\alpha}$ that for simplicity we assume to be energy- and spin-independent. The resulting  broadening of each level is given by 
$\Gamma_j=\Gamma_{jL}+\Gamma_{jR}$, with $\Gamma_{j\alpha}=\pi\rho (t^{j}_{\alpha})^2$.
Notation: We define $s_{i}={\rm sgn}{(t^{i}_{L}t^{i}_{R}t^{i+1}_{L}t^{i+1}_{R})}=\pm$.
For example, matrix elements of same sign result in  $s_{i}=+$, whereas one different sign yields $s_{i}=-$. We further define $s\equiv\{ s_{1} \cdots s_{N-1}\}$, and use $\gamma=\{\Gamma_{1L},\Gamma_{1R},\cdots,\Gamma_{NL},\Gamma_{NR}\}/\Gamma$, 
 with the mean level broadening
 $\Gamma=1/N\sum_{j}\Gamma_j$.
We assume  constant level spacing $\delta=\varepsilon_{di+1}-\varepsilon_{di}$. 
The local levels can be shifted in energy by a plunger gate voltage $V_g$, with 
 $\varepsilon_{dj}=j\delta -(V_g+V_{g0})$, where $V_{g0}=\frac{N-1}{2}\delta+\frac{2N-1}{2}U$.  This convention ensures that in case            of maximal symmetry ($t^{j}_{\alpha}=const.$ for all $j,\alpha$), the system possesses particle-hole symmetry at $V_g=0$.

\subsubsection{Transmission}
In the Appendix we generalize a result of Bruder, Fazio and Schoeller\cite{Bruder1996} to show
that the Aharonov-Bohm contribution to the linear conductance through the multi-terminal interferometer with open geometry with a multi-level quantum dot embedded in one arm  (see Fig.\ \ref{fig:AB} in the Appendix) can be expressed as
\begeq
	G^{AB}(T) = \frac{e^2}{h} |T_u| |t_d(T)| \cos(2\pi\Phi/\Phi_0 + \phi_0 + \alpha(T)).
\eeq
Here $T_{u}=|T_{u}|e^{{i\phi_{0}+i2\pi\Phi/\Phi_0}}$ is the energy- and temperature-independent transmission amplitude through the upper reference arm including the Aharonov-Bohm contribution $2\pi\Phi/\Phi_0$ to the phase, where $\Phi$ is the magnetic flux enclosed by the interferometer arms  and $\Phi_0=h/e$ is the flux quantum.
The equilibrium Fermi function of the leads are denoted by $f_0$.
The effective, temperature-dependent transmission amplitude  $t_{d}(T)$ through the lower arm including the quantum dot is given by
\begeq
   t_d(T)= \int dE 
   \left(-\frac{\partial f_0(E,T)}{\partial E}\right)
   T_d(E,T)
   \equiv |t_d|\ e^{i\alpha},
   \label{T}
\eeq
where
\begeq
   T_d(E,T) = \sum_{jj'}\sum_{\s\s'}
   2\pi\rho \ t^j_{L}   t^{j'}_{R}
   \GG^R_{j\s,j'\s'}(E,T).
   \label{Td}
\eeq
Therefore, 
only local properties like the local retarded Green's function $\GG^R_{j\s,j'\s'}$ 
and the Fermi function of the leads enter in the transmission amplitude through the quantum dot $t_d$ [Eq.\ (\ref{T})].
Thereby the local Green's function is evaluated for the model given in Eqs.\ (\ref{AM:Htot}) in equilibrium at temperature $T$.

In the zero temperature limit and in linear response, the dot produces purely elastic potential scattering between left and right leads, which can be fully characterized \cite{Pustilnik2001} by the eigenvalues $e^{i2\delta_\nu}$ ($\nu\hspace{-3pt}=\hspace{-3pt}a,b$) per spin of the $S$-matrix, and the transformation 
${\left( \begin{matrix} \cos\theta &\sin\theta \\ -\sin\theta&\cos\theta\end{matrix}\right)}$, 
that maps the left-right basis of lead operators onto the $a$-$b$ eigenbasis of $S$. The transmission amplitude through the dot then reads 
\begeq
	t_d=-i S_{LR}\hspace{-2pt} =\hspace{-2pt} \sin(2\theta) \sin(\delta_a-\delta_b) e^{i(\delta_a+\delta_b)},
	\label{td:n}
\eeq
where in general $\theta$ and $\delta_\nu$ are all $V_g$-dependent. 
The phase $\delta_\nu$ is related by the Friedel sum rule \cite{Langreth1966} to the charge (per spin) $n_{\nu}=\delta_\nu/\pi$ extracted by the dot from effective lead $\nu$. 
As $V_g$ is swept, 
the transmission amplitude goes through zero whenever $n_a=n_b \mbox{mod}1$, and a phase lapse by $\pi$ occurs.
Equation (\ref{td:n}) is useful for the special case of ``proportional couplings'',  $t^j_{L}=\pm\lambda t^j_R$ with $\lambda$ independent of $j$, in which the occupations $n_{a,b}$ take a simple form. Then the two effective leads $a$ and $b$ are the even and odd combinations of the left and right leads, respectively, with $\tan{\theta}=1/\lambda$ independent of $V_g$. Then each level either couples to the even \emph{or} the odd lead, and the occupations extracted from the leads are given by $n_{E,O}=\sum_{j \in {E,O}} n_{dj\s}$.
Note that if all levels are coupled to the same effective lead (which is the case for $s=\{+\cdots +\}$), the other effective lead decouples, thereby reducing the computational complexity\ significantly.

\subsubsection{The method}
We calculate the local Green's function $\GG^R$ needed for the transmission amplitude (Eqs.\ (\ref{T}) and (\ref{Td}), respectively) using the  numerical renormalization group method \cite{Krishna-murthy1980} (NRG), a well-established method for the study of strongly correlated impurity systems. For a review, see Ref.\ \onlinecite{Bulla2008}.
The key idea of NRG is the logarithmic discretization of the conduction band with a discretization parameter $\Lambda>1$. As a result, $H_{\rm l}$ is represented as a semi-infinite chain, where only the first site couples to the local level. The hopping matrix elements along the chain fall off exponentially like $\Lambda^{-(n-1)/2}$ with the site number $n$ (energy scale separation). The NRG  Hamiltonian can be solved iteratively by successively adding sites and solving the enlarged system, thereby increasing the energy resolution with each added site by a factor of $\Lambda^{1/2}$. The corresponding increase in Hilbert space is dealt with by a truncation strategy that keeps only the lowest $N_{\rm keep}$ states for the next iteration.

For the calculation of ${\rm Im}\mathcal{G}^R$ we use the full density matrix NRG \cite{Peters2006,Weichselbaum2007}, based on the only recently developed concept of a complete basis set within NRG \cite{Anders2005}. The real part of $\GG^{R}$ is obtained by Kramers-Kronig transformation.
Improvement of the results is obtained by the self-energy representation, where the $U$-dependent part of the impurity self-energy $\Sigma(\omega)=U\frac{F^R(\w)}{ G^R(\w)}$  is expressed by two correlation function \cite{Bulla1998}, which both are calculated with the full density matrix NRG.


\section{Results}
\label{sec:Results}

\begin{figure*}[t]
  \centering
  \includegraphics[width=0.8\textwidth,clip]{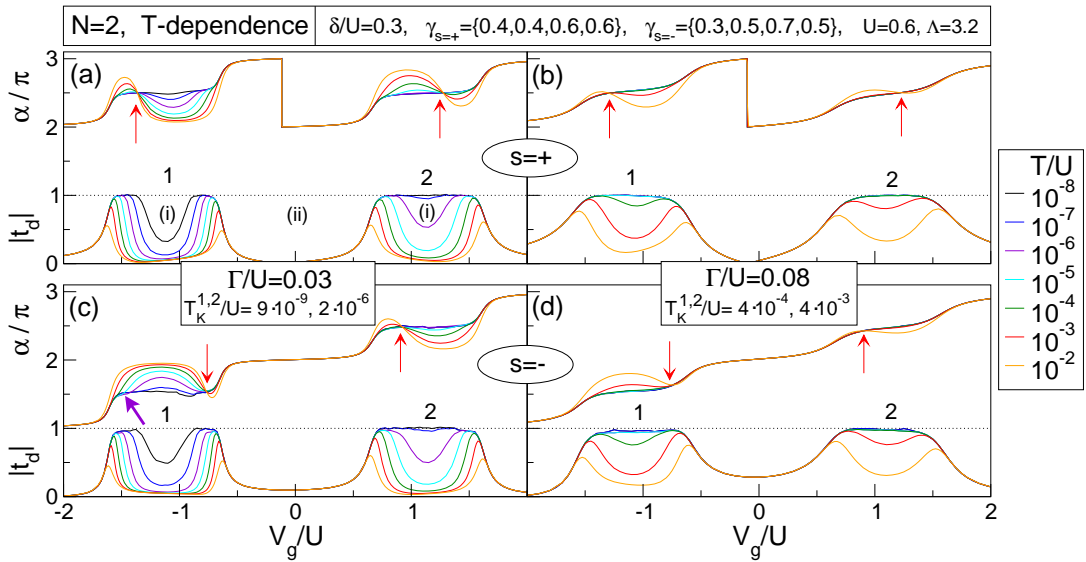}
  \caption{Transmission $t_d=|t_d| e^{i\alpha}$ through a spinful two-level quantum dot for various temperatures and constant couplings.
 Regimes (i), (ii), indicated in panel (a) only, refer to Kondo valleys or Coulomb blockade valleys, respectively (see text).
  The levels involved are indicated by their level number $1,2$.
  Level $2$ is coupled more strongly to the leads than level $1$,
  resulting in different bare Kondo temperatures, e.g.\ $T_K^{j=1}\hspace{-3pt}>\hspace{-3pt}T_K^{j=2}$. We use $\Gamma/U=0.03$ (a,c) and $\Gamma/U=0.08$ (b,d), thus $T_K^{(a,c)}\hspace{-3pt}<\hspace{-1pt}T_K^{(b,d)}$. The minimum value of the $T_K^j$ (in the center of the Kondo valleys) are indicated in the legends.
 In accordance with Ref.\ \onlinecite{Silvestrov2003}, we find shoulders in the phase (see e.g.\ the fat purple arrow and the purple curve  ($T/U=10^{-6}$) in (c) for level $1$) and an enhanced sensitivity of the phase to Kondo correlations compared to the magnitude, see e.g.\  the green ($T/U=10^{-4}$) curves in (d) for level $1$ or the purple curve ($T/U=10^{-6}$) for level $2$ in (a). There, the typical $\frac{\pi}{2}$-Kondo plateau in the phase is present, whereas the Kondo plateau in amplitude is not fully developed yet. 
 At certain points in gate voltage, say $V_g^{c_j}$ (as indicated by red arrows), we find stationary points where the curves for $\alpha$ for all temperatures intersect. The position of $V_g^{c_j}$ is shifted by the presence of a neighboring level, being repelled by or shifted towards the latter for $s=+$ or $-$, compare (a,c) or (b,d), respectively.
 Depending on the mesoscopic parameter $s=\pm$, the  phase either exhibits a sharp drop of $\pi$, accompanied by a zero in the amplitude $|t_d|$ ($s=+$, see (a,b)), or increases monotonically ($s=-$, see (c,d)) in the Coulomb blockade valleys.
  }
  \label{fig:2L:T}
\end{figure*}

In this Section we present our results for the phase and magnitude of the transmission amplitude $t_{d}$ through the quantum dot. The gate voltage $V_{g}$ is swept over a range sufficiently large that the full occupation spectrum of the quantum dot is covered ranging from $0$ to $2N$.
The exact distribution of the couplings seems to play only minor role for the transmission amplitude. Therefore we choose left-right symmetric coupling in the cases where all $s_j=+$, reducing the computational effort significantly, since then the odd channel decouples.
\\
In the regime of interest, the deep mesoscopic regime, the mean level spacing $\delta$ is much larger than the typical level widths $\Gamma_j$, $\delta/\G\gg 1$. Therefore electrons enter the dot one by one when increasing the gate voltage. Transport thus occurs mainly through one level at a time; more precisely, it occurs through a linear combination of all levels, where in the mesoscopic regime the level closest to the Fermi energy dominates.\cite{Karrasch2007a}
\\

The Section is organized as follows: We first eludicate the basic properties of the transmission amplitude for the example of a two-level system. Varying temperature $T$ (at fixed coupling $\Gamma$), or average coupling  $\Gamma$ (at fixed $T$), we study both possible choices  $s=+$ and $s=-$, respectively. In order to analyze the interplay of $s=+$ and $s=-$, we then present data for a three-level system for all four possible combinations of $s_{1},s_{2}$.
Additionally, this  has the advantage that for the middle level ``boundary effects'' (effecting the outermost levels) can be assumed to be  eliminated, thus 
the behavior of the middle level can be viewed as representative of a generic level in a multi-level quantum dot in the deep mesoscopic regime.


Unless otherwise noted, we use $U=0.6$. In order to cover all relevant energy scales with reasonable computational effort, we usually use $\Lambda=3.2$ for the two-level model and $\Lambda=3.5 $ in case of three levels. We checked that already by keeping $\sim 1000 $ states at each iteration, also for the two-channel calculations (that involve at least one $s_i=-$) the physical trends are captured qualitatively.
Note that since the eigenvalues of the scattering matrix are given by $e^{i2\delta_\nu}$, the transmission phase $\alpha$ is defined modulo $\pi$. For clarity of the Figures, curves showing $\alpha$ are shifted by multiples of $\pi$ as convenient.

\subsection{Two-level model}
\subsubsection{Temperature dependence}
\label{sec:2L:T}

%

Figure \ref{fig:2L:T} shows the transmission amplitude for both $s=+$ (a,b) and $s=-$ (c,d), for fixed dot parameters and various different temperatures. 
The mean level broadening is chosen to be $\Gamma/U=0.03$ in panels (a,c), and $\Gamma/U=0.08$ in panels (b,d). Therefore the ($V_g$-dependent) bare Kondo temperatures
\begeq
    T_K^j =\sqrt{\frac{\Gamma_j U}{2} }
   \exp{\left[-\pi\ \frac{\varepsilon_{dj}}{2U}\ \frac{(\varepsilon_{dj} +U)}{\Gamma_j}\right]}
	\label{TK}
 \eeq 
  vary in a lower-lying range of energies for panels (a,c) than for panels (b,d).
In all panels the relative coupling of the first and the second levels are chosen to be $\gamma=\{0.8,\,1.2\}$. Therefore, the bare Kondo temperature for level $1$ is lower than for level $2$, $T_K^{j=1}<T_K^{j=2}$, as indicated in the legends. The resulting difference in the temperature dependence can be nicely observed in the Figure. 
We  first describe those general properties of the transmission amplitude that qualitatively agree with those that one would obtain for just a single level, then discuss the effect of the presence of a second level.
\\

\emph{General properties:}
%
In the mesoscopic regime, where transport mainly occurs through one level at a time, 
two different regimes of transmission can be distinguished as $V_g$ is varied, as indicated in Fig.\ \ref{fig:2L:T}(a):
(ii) In the regime between the Kondo valleys, to be called ``Coulomb blockade valleys'', the transmission amplitude is mainly determined by the mesoscopic parameter $s$, showing a phase lapse only in case $s=+$, similar for both spinful and spinless models \cite{Karrasch2007a,Karrasch2007b}.

In the zero temperature limit, $T\ll T_K^{(j)}$, the transmission amplitude exhibits the typical Kondo behavior:
in the local-moment regime a typical  Kondo plateau forms, with $|t_{d}|$ approaching the unitary limit, $|t_{d}|\rightarrow 1$. In the mixed valence regime the magnitude changes rapidly as a function of $V_g$. In the Coulomb blockade valleys, transmission is suppressed by Coulomb interaction. 
The transmission phase increases by $\sim \pi /2 $ for each entering electron (see black curves for $\alpha$ in Fig.\ \ref{fig:2L:T}), increasing only slightly in between. In the Kondo valleys this results in a plateau at $\alpha\mbox{mod}\pi=\frac{\pi}{2}$, as direct consequence of the $\frac{\pi}{2}$ phase shift due to the formation of the Kondo singlet.

With increasing temperature, the Kondo effect is suppressed, thus the behavior in the middle of the Kondo valleys changes dramatically. The Kondo plateaus in $t_d$ and $\alpha$ disappear: 
The magnitude tends towards Coulomb blockade behavior, with a resonance of width $\sim\Gamma_j$ for each entering electron. 
The phase develops a $S$-like shape in the Kondo valleys with increasing temperature. 
As in the single-level case, 
all finite-temperature curves of the phase intersect the zero-temperature at the \emph{same} gate voltage, say $V_g^{c_j}$ (see red arrows).
We shall refer to this gate voltage as a ``stationary'' point (w.r.t.\  temperature).

As observed in the experiments of Ji \emph{et al.}\cite{Ji2000} and emphasized by Silvestrov and Imry,\cite{Silvestrov2003} the 
 transmission phase reacts more sensitively to the buildup of Kondo correlations with decreasing temperature than the transmission magnitude:
$\alpha$ approaches its $T=0$ behavior already at temperatures $T\simeq T_K$ (the $\frac{\pi}{2}$-plateau develops), whereas $|t_d|$ develops its plateau for $T$ significantly less than $T_K$ (see, the green curve ($T/U=10^{-4}$) for level $1$ in Fig.\ \ref{fig:2L:T}(b) or the purple curve ($T/U=10^{-6}$) for level $2$ in Fig.\ \ref{fig:2L:T}(a)).
Similar to the predictions of Silvestrov and Imry,\cite{Silvestrov2003} we find shoulders in the evolution of the phase, see for example the fat purple arrow and the purple curve ($T/U=10^{-6}$) in Fig.\ \ref{fig:2L:T}(c).
This indicates that the temperature is large enough to suppress  Kondo correlations in the deep local-moment regime (in the middle of the Kondo valley), where $T_K$ is very small. Towards the borders of the local-moment regime the crossover temperature for the onset of phase sensitivity increases (as does the Kondo temperature, see Eq.\ (\ref{TK})), eventually exceeding the temperature. Then the phase tends towards its zero-temperature behavior, thus producing shoulders.
\\

\begin{figure*}[t]
  \centering
  \includegraphics[width=0.8\textwidth]{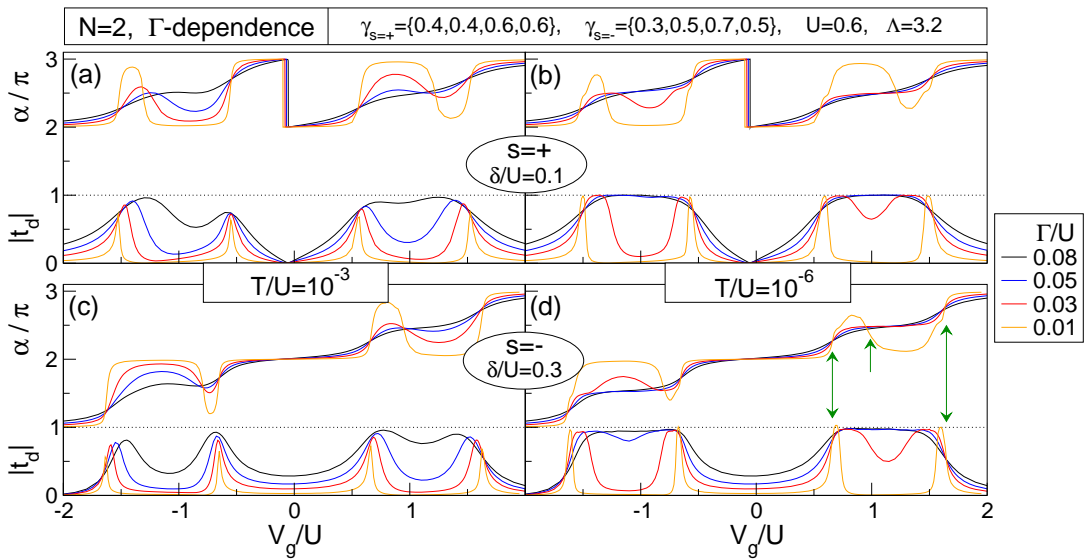}
  \caption{Transmission through a spinful two-level quantum dot for both choices of $s=\pm$ and
    various values of mean couplings $\Gamma$ at fixed temperature
    $T$, level spacing $\delta$ and relative couplings $\gamma$. Due to the mixing of the levels, no stationary points w.r.t.\ $\Gamma$ exist,
    $\Gamma$-independence exist, see text and the green arrows in (d).
  }
  \label{fig:2L:G}
\end{figure*}

\emph{Properties special to the multi-level model: }
The most obvious difference between the transmission amplitude of the many-level model in the mesoscopic regime compared to the single-level model is the phase behavior in the Coulomb blockade valleys between the levels.
Depending on $s$, i.e.\ on the relative sign of the tunnelling matrix elements of the two adjacent levels, the phase either exhibits a sharp drop (phase lapse) by $\pi$ in the $s=+$ case   (accompanied by a transmission zero, $|t_d|=0$), or evolves continuously for $s=-$ \cite{Oreg1997,Yeyati1995,Hackenbroich2001,Bruder1996,Karrasch2007a,Karrasch2007b}. 
Contrary to the non-monotonic phase evolution discussed above, this effect  occurs already at zero temperature and also exists for spinless models.\cite{Karrasch2007a,Karrasch2007b}   Therefore, the relevant energy scale for the temperature dependence of this phase lapse is not related to the Kondo temperature but to the level distance and width of the effective transport levels.\cite{Silvestrov2003} It is therefore not a relevant energy scale in the temperature range studied in this work. 

A further peculiarity for models with more than one level is the \emph{asymmetry} (w.r.t.\  the center of the Kondo valleys) of the transmission amplitude in the local-moment regime at finite temperature, introduced by the mixing of neighboring levels.
The asymmetry in phase can be characterized by the position of the stationary points, $V_g^{c_j}$ (indicted by red arrows in Fig.\ \ref{fig:2L:T}). In case $s=+$, these points are repelled by the neighboring level, whereas they are shifted to the latter for $s=-$, compare for example Fig.\ \ref{fig:2L:T}(a) and (c) or (b) and (d). %
For $\Gamma_1/\Gamma_2\neq 1$, the repulsion and attraction is enhanced or reduced compared to $\Gamma_1=\Gamma_2$ for the level that is coupled less or more strongly to the leads, respectively. Clearly, in the limit of one decoupled level (effective one-level system), the stationary point of the other level is symmetric w.r.t.\ the corresponding Kondo plateau.
The dips that form in the plateaus of the amplitude with increasing temperature develop a distinct asymmetry only for $T\gg T_{K}^{(j)}$, for which they  tend to shift towards the corresponding $V_g^{c_j}$. This is consistent with the fact that as the phase drop in the Kondo valley gets sharper with increasing temperature and approaches a quasi-phase lapse, the magnitude experiences a minimum, as for every complex function.
Interestingly, the asymmetry in phase is the same for all temperatures, thus already at temperature $T\lesssim T_K$ the phase ``knows'' 
in which direction (of $V_g$) the dip in magnitude will shift at higher temperatures.\\

\subsection{Dependence on the coupling strength}

In experiments, it is more convenient (and easier to control) to change the coupling strength between the quantum dot and the reservoirs than the temperature. Accordingly, Fig.\ \ref{fig:2L:G} presents the transmission amplitude for various values of $\Gamma$, keeping the temperature constant.
With decreasing $\Gamma$, the decrease of $T_K$ together with the suppression of Kondo  correlations is nicely illustrated. At fixed temperature $T>T_K$, the S-like shape of the phase evolution gets more pronounced and sharper with decreasing $\Gamma$.

In the single-level problem, in addition to stationary points w.r.t.\ temperature, we also find stationary points w.r.t.\ $\Gamma$ for $t_d$, i.e.\ for magnitude and phase of the transmission amplitude.
These occur at the outer flanks of the Kondo plateaus. 
Varying the mean coupling strength $\Gamma$ at fixed $\gamma$, $\delta$ and $T$ in the two-level model, as shown in Fig.\ \ref{fig:2L:G}, these points can still be recognized (indicated by green arrows in (d)), even though the $\Gamma$-independence is not perfect (within our numerical accuracy). 
We expect that due to the mixing of the levels, also the level distance $\delta$ has to be taken into account to recover these stationary points.
Between the levels, near $V_g/U\approx 0$, another stationary point seems to occur.

\begin{figure*}[t]
  \begin{center}
 \includegraphics[width=0.9\textwidth,clip]{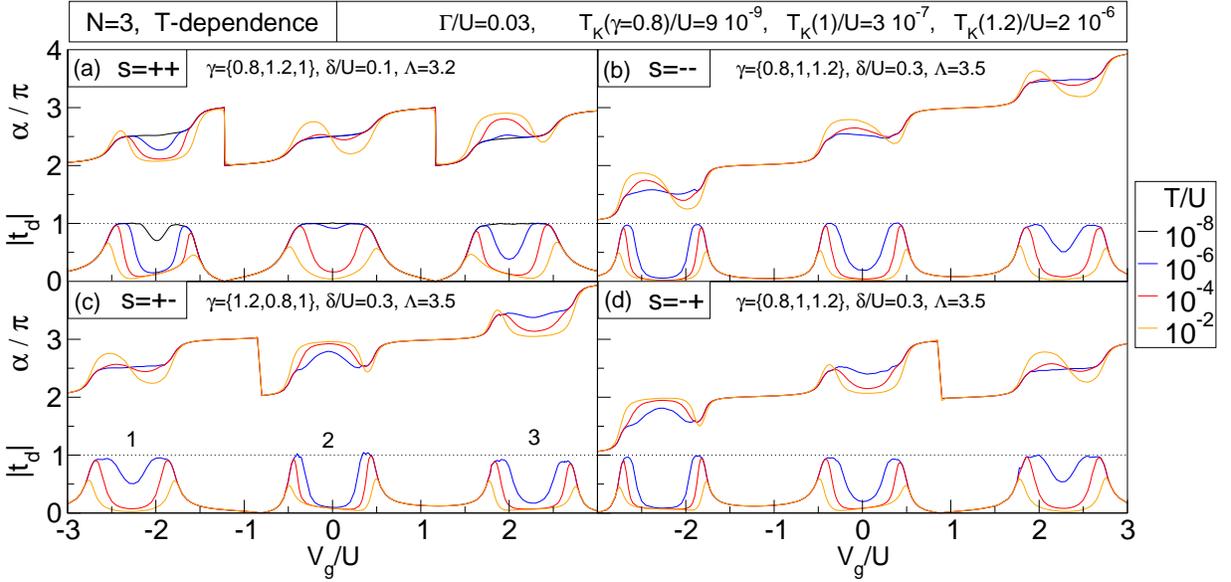}
 \end{center}
  \caption{Transmission through a three-level model for all four possible combinations of $s=s_1,s_2$ and various temperatures, for fixed $\Gamma/U=0.03$.
The two-channel calculations for (b,c,d) qualitatively capture the physical trends. 
    The asymmetry in the Kondo valleys is determined by both $s$ and $\gamma$. For convenience the figure legends for $\gamma$ display only the total relative coupling of each level. The minimal bare Kondo temperatures are indicated.
    (a) $s=++$: $\gamma= \left\{0.6,0.6,0.4,0.4,0.5,0.5\right\}$. The case $T/U=10^{-8}$ is included only for this panel.
    (b) $s=--$: $\gamma= \left\{0.5,0.3,0.4,0.6,0.5,0.7\right\}$.
    (c) $s=+-$: $\gamma= \left\{0.4,0.8,0.4,0.4,0.3,0.7\right\}$.
    (d) $s=-+$: $\gamma= \left\{0.5,0.3,0.3,0.7,0.7,0.5\right\}$.
  }
  \label{fig:3L}
\end{figure*}

\subsection{Three-level model}

Naturally, the question arises about the effects of several levels, with different choices of 
$s_i=\pm$, which is present only for models with more than two
levels. Assuming that in the mesoscopic regime only neighboring levels
mix significantly, i.e.\ simultaneously influence transport, any local
level of a quantum dot (except the lowest or highest one) can be
represented adequately by the middle level of a three-level model.

In Fig.\ \ref{fig:3L} we present numerical data of a three-level model
for all four possible combinations of $s=s_1,s_2$ and various
temperatures. The second level is influenced by the effect of both
$s_1$ and $s_2$, resulting in an effective enhancement or compensation
of the asymmetry of the stationary point $V_g^{c_2}$ of level $2$,
as discussed in Section \ref{sec:2L:T}. Also the relative strength of
the level-couplings (given by $\gamma$) has to be considered.
In Fig.\ \ref{fig:3L}(a), both $s$ and $\gamma$ symmetrize the transmission curves
of the middle level, whereas in panel \ref{fig:3L}(b) $\gamma$ shifts $V_g^{c_2}$
to positive $V_g$.  In panels \ref{fig:3L}(c) and \ref{fig:3L}(d) both $s$ and $\gamma$ tend
to increase the asymmetry.
\\

Therefore, the transmission phase through a spinful quantum dot with
Kondo correlations present has $S$-like shape in the local-moment
regimes at $T\gg T_K$. Analogously to experiments, we find an
asymmetry of this $S$-like shape. It is determined by both the relative
strength $\gamma$ and the sign $s$ of the level couplings.

\section{Conclusion}
\label{sec:Conclusion}

In this paper we present temperature-dependent NRG calculations of the magnitude and phase of the transmission amplitude through a multi-level quantum dot in the regime $\delta/\Gamma\ll 1$. Clearly, the Kondo correlations are suppressed with increasing temperature. The presence of neighboring levels results in a $V_g$-asymmetry in the finite temperature modulation of the Kondo valleys. The asymmetry depends on the relative signs of the tunneling matrix elements as well as on the relative couplings of the adjacent levels. Further, sharp phase lapses may occur between the levels. Studying a three-level model, the middle level 
can be understood as a representative of a generic level in a multi-level quantum dot.

Throughout the paper, we deliberately focussed
 only on the deep mesoscopic regime,
for which the results can be understood rather straightforwardly. The crossover into the regime $\delta / \Gamma \simeq 1$,
which is certainly of interest too in order to understand 
the fate of Kondo physics in the universal regime, and which we believe to be the regime relevant for the experiments of Ji \emph{et al.},\cite{Ji2000,Ji2002} will 
be left as a subject for future studies.

\section{Acknowledgement} %
 We acknowledge helpful discussions with Moty Heiblum, Michele Zaffalon, Vitaly Golovach and Michael Pustilnik.  This research was supported by the DFG
  through De-730/3-2, SFB631 and SFB-TR12, and by
  DIP-H.2.1.  Financial support of the German Excellence Initiative
  via the ``Nanosystems Initiative Munich (NIM)'' is gratefully
  acknowledged.

\appendix
\section{ Conductance formula for multi-terminal geometry}

\subsection{General case}
We generalize the current formula derived in by Bruder, Fazio and Schoeller\cite{Bruder1996} for a
single-level quantum dot embedded into one arm of an Aharonov-Bohm
interferometer with two-terminal geometry to a 
multi-terminal geometry with a multi-level dot (as used in the Heiblum group \cite{Schuster1997,Ji2000,Ji2002,Avinun-Kalish2005,Zaffalon2007}).
\\

\begin{figure}[ht]
  \centering
 \includegraphics*[width=0.8\columnwidth]{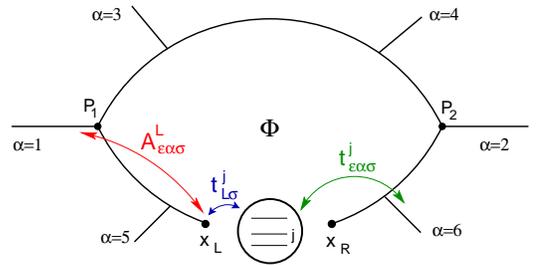}
  \caption{Geometry of the multi-terminal  Aharonov-Bohm interferometer with a multi-level quantum dot embedded in the lower arm. The different tunnelling amplitudes used in the text are indicated. $\Phi$ is the magnetic flux penetrating the interferometer. }
  \label{fig:AB}
\end{figure}

Consider a $N$-level quantum dot described by $H_{\rm d}$
[Eq.\ (\ref{AM:H_imp})] embedded in one arm of an Aharonov-Bohm
interferometer connected to $M$ leads, as depicted in Fig.\
\ref{fig:AB}. 
Each lead, and each arm connecting them, is assumed to
support only a mode. The tunnelling between the local levels
$j=1\cdots N$ on the quantum dot and the leads $\alpha=1\cdots M$ is
described by \begeq H_{\rm t} = \sum_{j \s} \sum_{\alpha k } t^j_{\eps
  \alpha\s} c^\dag_{\alpha \eps \s}d_{j\s} + \mbox{H.c.}
   \label{APP:HT}
\eeq
Here $t^j_{\eps \alpha\s}=\sum_{i=L,R} t^j_{i \s} A^i_{\eps \alpha \s}$ (indicated in green in Fig.\ \ref{fig:AB}) is chosen real, 
where 
$t^j_{i \s}=\langle x_i|j\s\rangle$ (blue) is the amplitude to get from dot state 
$|j\s\rangle$ of level $j$ and spin $\s$ 
to point $x_i$  on side $i=L,R$ of the dot,
and 
$A^i_{\eps \alpha \s}=\langle \eps \alpha\s|x_i\rangle$ (red)
is the amplitude to get from point $x_i$  to lead state $|\eps \alpha\s\rangle$ in lead $\alpha$ with energy $\eps$ and spin $\s$, see Fig.\ \ref{fig:AB}.

Following B\"uttiker,\cite{Buttiker1992}  
the current operator in reservoir $\alpha$ is given by 
\begeq
   \hat{I}_\alpha (t)
   =
   \frac{e}{h}
   \sum_{\eps\eps'}   \sum_{\s\s'}
   \frac{1}{\rho}
   \left[c^\dag_{\eps'\alpha\s'}(t) c_{\eps\alpha\s}(t)
     -  b^\dag_{\eps'\alpha\s'}(t) b_{\eps\alpha\s}(t)
     \right],
     \label{APP:I}
\eeq
where $\rho$, the density of states, is assumed to be
constant and equal for each reservoir.  The first term inside the
bracket stands for the incident, the second term for the reflected
current in reservoir $\alpha$, thus $b_{\eps\alpha\s}=\sum_\beta
S^\eps_{\alpha\beta} c_{\eps\beta\s}$, with $S^\eps_{\alpha\beta}$ the
scattering amplitude to get from lead $\beta$ to lead $\alpha$ with
energy $\eps$.  Defining the lesser, retarded and advanced correlation
functions 
\begeqn 
\GG^<_{\mu, \mu'}(t-t') &\equiv& \frac{i}{\hbar} \langle
a^\dag_{\mu'}(t') a_{\mu}(t) \rangle \nonumber
\\
&=& 
\int \frac{dE}{2\pi\hbar}\ e^{-iE(t-t')/\hbar}\ \GG^<_{\mu,\mu'}(E),
\\
\GG^{R,A}_{\mu,\mu'}(t-t') &\equiv& 
-\frac{i}{\hbar} \theta(\pm(t-t'))
\langle \left[ a^\dag_{\mu'}(t'), a_{\mu}(t)\right]_+ \rangle \nonumber
\\
&=& \int \frac{dE}{2\pi\hbar}\ e^{-iE(t-t')/\hbar}\
\GG^{R,A}_{\mu,\mu'}(E), \eeqn where $a_\mu$ denotes a fermionic operator
with composite index $\mu$, the expectation value of the current operator
(\ref{APP:I}) can be expressed as 
\begeqn \langle \hat{I}_\alpha (t)\rangle &=&
\frac{e}{h} \sum_{\eps\eps'}\sum_{\beta\beta'}\sum_{\s\s'}
\frac{1}{\rho} \left[\delta_{\alpha\beta'}\delta_{\alpha\beta} -
  S^{\star\eps'}_{\alpha\beta'}S^{\eps}_{\alpha\beta} \right]
\nonumber
\\
&&\times(-i) \int \frac{dE}{2\pi}\
\GG^<_{\eps\beta\s,\eps'\beta'\s'}(E).
   \label{APP:Imean}
\eeqn

To calculate $\GG^<(E)$ in Eq.\ (\ref{APP:Imean}), we use the standard
Dyson equation for the Keldysh $2 \times 2$ matrix Green's function\cite{Rammer1986}
$\GGh (E)$, 
\begeqn
\GGh_{\eps\alpha\s,\eps'\alpha'\s'}(E) &=&
\delta_{\eps\eps'}\delta_{\alpha\alpha'}\delta_{\s\s'}
\GGh^0_{\eps\alpha\s}(E)
      \label{APP:Dyson}
      \\
   &&
   \hspace{-70pt}
   +\sum_{jj'}
   \GGh^0_{\eps\alpha\s}(E) 
   t^j_{\eps\alpha\s}  \GGh^d_{j\s,j'\s'}(E)  t^{\star j'}_{\eps'\alpha'\s'}
   \GGh^0_{\eps'\alpha'\s'}(E),
\nonumber
\eeqn
which yields
\begeqn
   \GG^<_{\eps\alpha\s,\eps'\alpha'\s'}(E)
   &=&
   \delta_{\eps\eps'}\delta_{\alpha\alpha'}\delta_{\s\s'} 
   \GG^{0<}_{\eps\alpha\s}(E)
      \label{APP:Glessa1}
      \\
      &+& \sum_{jj'} t^j_{\eps\alpha\s} 
      \left[\hspace{-2pt}
        A\hspace{-2pt}+\hspace{-2pt}B\hspace{-2pt}+\hspace{-2pt}C\right] t^{\star j'}_{\eps'\alpha'\s'}\; ,
      \nonumber \eeqn 
where terms in square brackets are given by
\begin{eqnarray*}
   A
   &= &
   \GG^{0R}_{\eps\alpha\s}(E) \ 
   \GG^R_{j\s,j'\s'}(E) \ 
   \GG^{0<}_{\eps'\alpha'\s'}(E) \; ,
   \\
   B
   &=&
   \GG^{0R}_{\eps\alpha\s}(E) \ 
   \GG^<_{j\s,j'\s'}(E) \ 
   \GG^{0A}_{\eps'\alpha'\s'}(E) \; ,
   \\
   C
   &=&
   \GG^{0<}_{\eps\alpha\s}(E) \ 
   \GG^A_{j\s,j'\s'}(E) \ 
   \GG^{0A}_{\eps'\alpha'\s'}(E) \; ,
\end{eqnarray*}
where the free Green's functions for the leads have the form
\begeqn
   \GG^{0R,A}_{\eps\alpha\s}(E)&=& \frac{1}{E-\eps\pm io^+},
   \\
   \GG^{0<}_{\eps\alpha\s}(E) &=& 2\pi i  f_{\alpha}(E)  \delta(\eps-E),
\eeqn
with $f_{\alpha}(E)$ the Fermi function of lead $\alpha$.
Inserting $\GG^<$ [Eq.~(\ref{APP:Glessa1})] into Eq.\
(\ref{APP:Imean}), the current can be written as \begeq \langle
\hat{I}_\alpha\rangle = I_\alpha^0 + \delta I_\alpha.
   \label{APP:I0_dI}
   \eeq $I_\alpha^0$ arises from the first term of Eq.\
   (\ref{APP:Glessa1}). It describes the situation when the dot is
   completely decoupled ($t^j_{\eps\alpha\s}=0$), thus does not
   contribute to Aharonov-Bohm oscillations.  The influence of the
   quantum dot on the Aharonov-Bohm oscillations is caused by $\delta
   I_\alpha$, arising from the second expression of Eq.\
   (\ref{APP:Glessa1}).  Performing the energy sums $\sum_{\eps\eps'}$
   in Eq.\ (\ref{APP:I0_dI}) or (\ref{APP:Imean}), respectively, the
   two contributions to the current read 
\begin{subequations}
\label{subeq:currentresult}
\begeqn
I_\alpha^0  & = &
   \frac{e}{h}\int dE \sum_\s \sum_\beta \left[\delta_{\alpha\beta}-
     |S_{\alpha\beta}^E|^2\right] \! f_\beta (E) \; , \qquad 
\phantom{.}
\\
   \delta I_\alpha
   & = & 
   \frac{e}{h}  {\rm Re }   \bigg\{ 
   \int\hspace{-3pt} dE \hspace{-2pt}
   \sum_{\beta\beta'}\sum_{jj'}\sum_{\s\s'}\hspace{-1pt}
   \left[\delta_{\alpha\beta'}\delta_{\alpha\beta}
     - S^{\star E}_{\alpha\beta'}S^{E}_{\alpha\beta}
   \right] \biggr.
   \nonumber
\\
	&& \hspace{-12pt}
	\times\
  	\pi\rho \ t^j_{E\beta\s}   t^{\star j'}_{E\beta'\s'}
	(-i)
           \label{APP:BFS}
   \\
&&\hspace{-12pt}
 \biggl. \times   \left[
     2 \GG^R_{j\s,j'\s'}(E) f_{\beta'}(E) 
     \hspace{-1pt}+\hspace{-1pt}
        \GG^<_{j\s,j'\s'}(E) \right]\hspace{-1pt}\bigg\}.
\nonumber
\eeqn
\end{subequations}


\subsection{Simplification to effective 2-lead geometry}

For the experimental setup used by Schuster \emph{et al.}\cite{Schuster1997} (and equivalently for the ensuing papers \cite{Ji2000,Ji2002,Avinun-Kalish2005,Zaffalon2007})
to measure transmission phase shifts, two simplifying assumptions can
be made. The first allows us to neglect non equilibrium effects, the
second to perform NRG calculations for a simplified geometry,  in which
the dot is coupled only to two leads.

\emph{ (i) Neglect of non equilibrium effects:} 
In the experimental
setup used by Schuster \emph{et al.},\cite{Schuster1997} the leads
$\alpha = 3, 4, 5$ and 6 serve as draining reservoirs (to prevent
multiple traversals of the ring, see below), and are all kept at
chemical potential $\mu_\alpha = 0$. This also fixes the chemical
potential of the ring, referred to as ``base region'' in Ref.\ 
\onlinecite{Schuster1997}, to equal zero.  Lead 1 and 2 serve as emitter and
collector, respectively, with chemical potentials $\mu_1$ and $\mu_2$,
and Fermi functions $f_{1,2} (E)=f_0(E - \mu_{1,2})$.  Now, the point
contacts between emitter or collector and the base region (marked
$P_1$ and $P_2$ in Fig.~\ref{fig:AB}) are so small that the voltage drops occur directly at
these point contacts, and \emph{not} at the tunnel barriers coupling
the dot to the ring. Thus, while the emitter or collector inject or
extract electrons into or from the base region, respectively, this is
assumed to happen at a sufficiently small rate that the base region is
not disturbed. In other words, \emph{we may assume that the dot, ring,
  and electrodes 3, 4, 5, 6 are all in equilibrium with each other,}
and that the dot Green's functions $ \GG^{R,A,<}_{j\s,j'\s'}(E)$ do
not depend on $\mu_1$ and $\mu_2$ at all. Thus, the lesser function
can be expressed in terms of the retarded and advanced ones using the
following standard equilibrium relation: \begeq \GG^<_{j\s,j'\s'}(E) =
-f_0(E) \left[ \GG^R_{j\s,j'\s'}(E)-\GG^A_{j\s,j'\s'}(E)\right] \; .
\eeq The conductance in the linear response regime can be obtained by
taking $\mu_1 - \mu_2 = eV$, where $e=|e|$, e.g.\ by setting 
\begeq
	\mu_{1} =0,\ \ \ \ \ \ \ \mu_2=- eV,
	\label{APP:mu}
\eeq
and calculating $G=\partial I_1/\partial V$, with $I_1$ given by
Eq.~(\ref{subeq:currentresult}).

\emph{ (ii) Reduction to two-lead geometry:} The reason why a
multi-lead geometry was used in experiment is to avoid phase-rigidity:
in an Aharonov-Bohm ring connected to only two leads, the transmission
phase of the dot does not vary smoothly with gate voltage, but can
assume only two distinct values, differing by $\pi$.  A multi-lead
geometry avoids this by strongly reducing the probability amplitudes
for paths from emitter to collector to traverse the ring multiple
times, since with each traversal of the ring the probability increases
that electrons travelling in the ring are ``siphoned off'' into the
side arms. We shall exploit this fact by making the assumption that
\emph{the probability amplitudes for multiple traversals of the ring
  are negligibly small}.  This assumption allows us to replace the
multi-lead geometry with one where the Aharonov-Bohm ring is coupled
to only two leads, i.e.\ $\alpha$ is restricted to the values 1 and 2
(corresponding to emitter and collector), while multiple traversals of
the ring are eliminated (by hand) by the following specification: The
amplitude $t_{\eps\alpha\s}^j$ to get from state $|j\s\rangle$ on the
dot to state $|\eps\alpha\s\rangle$ in lead $\alpha$ is taken to be
nonzero only for the \emph{short, direct} path from the dot to lead
$\alpha$, \emph{without} traversing the upper arm (more correctly: we take $A^L_{\eps\alpha\s}=0$ for $\alpha=2,4,6$ and $A^R_{\eps\alpha\s}=0$ for $\alpha=1,3,5$). 
 When calculating
the current we do allow for direct paths from lead 1 to 2 via the
upper arm, and lump all flux-dependence into the corresponding
scattering amplitude, taking $S^{E}_{12} \sim e^{i2\pi \Phi/\Phi_0}$.
However, the upper arm is \emph{ignored} for the calculation of the
equilibrium local retarded or advanced Green functions
$\GG^{R,A}_{j\s,j'\s'}(E)$ using NRG.  For the latter purpose, we thus
use a model of a multi-level dot coupled to two independent leads, say
$L$ and $R$, with \emph{equal} chemical potentials $\mu_L = \mu_R$,
representing the two segments of the ring to the left and right of the
ring, coupled to it by tunnelling contacts. (These two segments should
be treated as independent leads, due to the assumption of no multiple
traversals made above.)
With the assumptions (i) and (ii) just described, let us now obtain an
expression for that part of the conductance showing Aharonov-Bohm
oscillations with applied flux, 
$ G^{AB} =\frac{\partial I^{AB}_1}{\partial V}$, where
$I_1^{AB}$ is that part of the current in lead 1 
depending on $ e^{i2\pi \Phi/\Phi_0}$.
For the chemical potentials given by Eq.\ (\ref{APP:mu}), this corresponds to evaluate
Eq.~(\ref{APP:BFS}) with $\alpha=1$, $\beta'=2$ and $\beta=1$, and
we readily obtain 
\begeq 
G^{AB}(T)  = 
\frac{e^2}{h} \int dE \ {\rm Re }\ \left[T_u^\star(E) T_d(E) \right]
\left(-\frac{\partial f_0(E)}{\partial E}\right) \; ,  
\eeq 
where 
\begeqn
\hspace{-4pt} T_d(E) & = & 
\sum_{jj'}\sum_{\s\s'} 2\pi\rho \ t^j_{E1\s}
\GG^R_{j\s,j'\s'} (E) t^{\star j'}_{E2\s'}
   \label{APP:Td}
\\
  \hspace{-4pt}  T_u^\star(E) & = &i S^{\star E}_{12}S^{E}_{11}\hspace{-2pt} =\hspace{-2pt} 
|T_u(E)| e^{i(2\pi \Phi /\Phi_0 + \phi_0(E))}
\eeqn
may be interpreted as the transmission amplitudes through the lower
and upper arms, respectively.

Assuming the transmission amplitude $T_u$ through the upper arm to be
energy- and temperature-independent, 
the Aharonov-Bohm contribution to the conductance is
given by 
\begeqn 
\hspace{-4pt} G^{AB}(T) = \frac{e^2}{h}\hspace{-3pt} |T_u| |t_d(T)|\hspace{-1pt} \cos(2\pi\Phi/\Phi_0\hspace{-2pt} + \hspace{-2pt} \phi_0 \hspace{-2pt} +\hspace{-2pt} 
\alpha(T)).
   \label{APP:GAB}
\eeqn
Then, the temperature-dependent
\emph{magnitude and phase} of the transmission amplitude through the quantum
dot, 
\begeq 
t_d(T)= \int dE \left(-\frac{\partial f_0(E,T)}{\partial
    E}\right) T_d(E,T)\equiv |t_d(T)| e^{i\alpha(T)},
   \label{APP:T}
\eeq
can be (i) extracted via Eq.\ (\ref{APP:GAB}) from the experimental results as well as (ii) calculated with NRG using Eq.\ (\ref{APP:Td}).


\end{document}